\documentclass[lettersize,journal]{IEEEtran}
\usepackage{amsmath,amsfonts}
\usepackage{algorithmic}
\usepackage{algorithm}
\usepackage{array}
\usepackage[caption=false,font=normalsize,labelfont=sf,textfont=sf]{subfig}
\usepackage{textcomp}
\usepackage{stfloats}
\usepackage{url}
\usepackage{verbatim}
\usepackage{graphicx}
\usepackage{color}
\usepackage{cite}

\IEEEoverridecommandlockouts

\title{\LARGE \bf
A Mechanomyography-Based Integrated Sensing–Stimulation System for Fatigue-Aware Neuromuscular Therapy
}

\author{
Xiao Jin$^{1}$, Cong Liu$^{2}$, Shibo Jing$^{3}$, Zhenhua Yu*$^{1}$$^{2}$\\
\thanks{*Corresponding author: Zhenhua Yu}%
\thanks{This work was partly supported by Capital Equipment funding of University of Aberdeen.}%
\thanks{$^{1}$Department of Computer Science, University of Aberdeen, AB24 3UE, Aberdeen, UK.
       }%
\thanks{$^{2}$Dyson School of Design Engineering, Imperial College London, SW7 2DB, UK.
        {\tt\small e-mail: zhenhua.yu@abdn.ac.uk}}%
\thanks{$^{3}$Department of Mechanical Engineering, Imperial College London, SW7 2DB, London, UK.
        }%
}
\begin{document}
\maketitle

\begin{abstract}
Muscle fatigue limits the effectiveness of functional electrical stimulation (FES), yet fatigue monitoring during active stimulation remains challenging because electromyography (EMG) is severely contaminated by stimulation artifacts. To address this issue, this study proposes an integrated mechanomyography-based sensing framework (MMG-FES) for artifact-resistant fatigue monitoring during stimulation. Fatigue states were first characterized during voluntary movement using established EMG spectral indicators, including median frequency (MF) and mean power frequency (MPF), and then linked to MMG-derived features, including high-frequency energy ratio (HFER) and spectral bandwidth (BW), for fatigue recognition under stimulation conditions. Experimental results showed consistent fatigue-related trends, with EMG spectral features decreasing and MMG features exhibiting coordinated spectral redistribution across repeated contractions. Using these features, the proposed framework achieved fatigue-state recognition accuracies of 0.93 and 0.96 with logistic regression and random forest, respectively. These findings support MMG as a physiologically meaningful and stimulation-compatible modality for fatigue-aware closed-loop neuromuscular stimulation.
\end{abstract}


\section{Introduction}
\label{introduction}

Functional electrical stimulation (FES) has become an essential technique in modern rehabilitation engineering, offering an effective means to restore or assist motor functions in individuals with impaired muscle control. By delivering controlled electrical impulses to peripheral nerves or muscles, FES induces muscle contractions that can compensate for voluntary motor deficits caused by conditions such as stroke, spinal cord injury (SCI), or neuromuscular disorders~\cite{an, POPOVIC2014795}. Over the past decade, FES has evolved from a laboratory concept to a clinically validated intervention, enabling improvements in gait, grasping, and postural control~\cite{ragnarsson2008fes, Bellman2016FEScycling}. According to recent reports, more than 15 million people worldwide suffer from motor dysfunctions that could potentially benefit from FES-assisted therapies, while the global FES device market is projected to exceed USD~1.2~billion by~2030, driven by the growing demand for personalized neurorehabilitation~\cite{who2022stroke, globalmarket2023fes}. 

Despite these advances, the effectiveness of FES critically depends on the precise control of stimulation parameters—such as current amplitude, pulse width, and frequency—to achieve stable and fatigue-resistant muscle responses~\cite{merletti2001electrical, liberson1961functional}. However, conventional open-loop stimulation strategies often fail to account for dynamic muscle conditions during long-term use, leading to rapid fatigue, discomfort, and limited functional recovery~\cite{downey2018towards}. These challenges highlight the necessity of developing closed-loop FES systems capable of continuously monitoring muscle state and adjusting stimulation in real time to maintain optimal performance and comfort~\cite{li2018realtime}.

To achieve effective and adaptive control of FES, accurate monitoring of muscle activity is essential. Various biosignal modalities have been employed to quantify muscle state, including electromyography (EMG), force myography (FMG), and sonomyography (SMG)~\cite{xia2020review, yang2021wearable}. Among these, EMG remains the most widely adopted technique for assessing neuromuscular activation by measuring electrical potentials generated during muscle contraction~\cite{de2009surface}. However, EMG-based control faces significant limitations under electrical stimulation conditions. The external stimulation pulses often introduce severe artifacts that obscure the underlying voluntary EMG signals, preventing real-time decoding of muscle activity~\cite{mandrile2003stimulation}. Although several artifact removal algorithms have been proposed, they typically rely on offline post-processing or introduce substantial latency, which restricts their use in closed-loop FES systems~\cite{khan2024eeg}.

In contrast to electromyography (EMG), which measures electrical potentials generated during muscle activation, mechanomyography (MMG) detects the low-frequency mechanical oscillations produced by muscle fiber contractions, typically within the 5–100~Hz range~\cite{barroso2013measurement, orizio1993muscle}. These oscillations reflect both the onset of motor unit recruitment and the mechanical response of the musculotendon system, providing a direct mechanical correlate of muscle function\cite{Rohlén2020A}. MMG signals have been shown to correlate strongly with muscle force~\cite{xiao2025multisensor}, fatigue, and firing rate~\cite{liu2025continuous,uwamahoro2021assessment}. More importantly, because MMG measures mechanical rather than electrical activity, it is inherently immune to the stimulation artifacts that often corrupt EMG recordings during functional electrical stimulation (FES). Recent studies have demonstrated that MMG can continuously track muscle fatigue and contraction dynamics in real time under FES conditions, enabling accurate estimation of neuromuscular state even during ongoing stimulation~\cite{Ha}. This capability makes MMG a promising sensing modality for closed-loop FES systems that require artifact-free, real-time monitoring and adaptive control of muscle performance.

Recent studies have also shown that miniaturized microphone-based MMG sensors can achieve high sensitivity while maintaining comfort and flexibility for wearable integration. Capacitive MEMS microphones (CMS), initially designed for acoustic measurements, have demonstrated superior signal-to-noise ratios and the ability to detect muscle deformation with sub-millimeter precision~\cite{mi}. For instance, Yoshimi \emph{et al.} demonstrated that condenser microphone arrays could precisely capture muscle vibration waveforms corresponding to contraction intensity~\cite{kim2008forcetremor}, while Liu \emph{et al.} achieved continuous fatigue estimation during FES using MEMS-based MMG sensors coupled with predictive modeling~\cite{liu2025continuous}. These findings highlight the potential of CMS-based MMG sensors as compact, low-cost, and artifact-resistant platforms for muscle monitoring.

However, despite the rapid progress in both MMG sensing and FES control, the two domains remain largely decoupled. Current systems typically record MMG and deliver FES through independent channels or in alternating temporal modes to minimize interference~\cite{Gon}. To overcome the conventional separation between neuromuscular sensing and stimulation, this study presents an integrated capacitive MEMS (CMS)–based MMG array that enables simultaneous muscle vibration sensing and electrical neuromodulation within a single flexible substrate. Unlike prior systems using discrete modules~\cite{downey2018towards}, the platform embeds multiple CMS sensors in a spatial array to continuously map muscle mechanical activity during active stimulation. Rather than targeting joint motion, this work treats electrically evoked muscle vibration as a therapeutic mechanical stimulus—analogous to deep-tissue vibration—aimed at modulating muscle state and alleviating fatigue. Notably, CMS-based MMG signals remain robust during stimulation, allowing reliable observation where EMG is often corrupted.

To enable reliable sensing during electrical stimulation, the proposed system integrates MMG directly within the FES process to capture the mechanical response of muscle tissue in real time. The CMS-based MMG array provides stable, artifact-resistant monitoring of contraction dynamics, overcoming the limitations of conventional approaches that require temporal separation between sensing and stimulation. By embedding sensing and actuation within a unified framework, this approach supports continuous neuromuscular state estimation and facilitates the development of closed-loop FES systems with adaptive control capability. Clinical evidence further indicates that electrically or mechanically induced muscle activation can reduce spasticity, enhance strength~\cite{Maghbouli2021The}, accelerate motor recovery~\cite{Percival2021Local}, and improve rehabilitation outcomes~\cite{Xu2024Effect}, highlighting the importance of reliable, continuous monitoring during FES.

Building on this foundation, we propose an MMG-based functional electrical stimulation (MMG-FES) framework in which MMG-derived features are used to continuously characterize stimulation-induced muscle mechanical responses during ongoing stimulation. By quantifying vibration intensity and spectral characteristics over time, the system enables real-time assessment of muscle state under stimulation-on conditions, including fatigue accumulation and recovery-related changes. This work demonstrates how mechanically grounded MMG feedback can support fatigue-aware adjustment of stimulation usage, advancing neuromuscular therapy from open-loop, parameter-driven operation toward a muscle-state–aware intervention paradigm with implications for adaptive and personalized rehabilitation technologies~\cite{jing2025enhancing,11039026}.

The main contributions of this work are summarized as follows:
\begin{itemize}
\item We propose a unified MMG--FES sensing and stimulation framework that enables artifact-free neuromuscular monitoring during active electrical stimulation, addressing the fundamental limitation of EMG in FES scenarios.

\item We identify fatigue-related mechanical signatures from MMG signals and establish a physiologically grounded fatigue representation by linking MMG spectral features with established EMG-derived fatigue indicators, enabling reliable fatigue characterization under stimulation conditions.

\item We develop and validate a cross-condition fatigue recognition strategy that transfers fatigue knowledge from voluntary contractions to stimulation-evoked responses, demonstrating the potential of the proposed approach for real-time fatigue-aware adaptive electrical stimulation across multiple muscle groups.
\end{itemize}

\section{Sensor Architecture and Characterization}
\label{Hard}

Approved by the University of Aberdeen Ethics Committee (ID: 8891100), experiments involved repeated daily lower-limb tasks in a controlled lab. All channels were sampled at 1440 Hz, with data collected over two weeks to minimize fatigue effects.

\subsection{Design of the MMG-FES Integrated Sensor Module}

\begin{figure}[ht]
    \centering
    \includegraphics[width=0.9\linewidth]{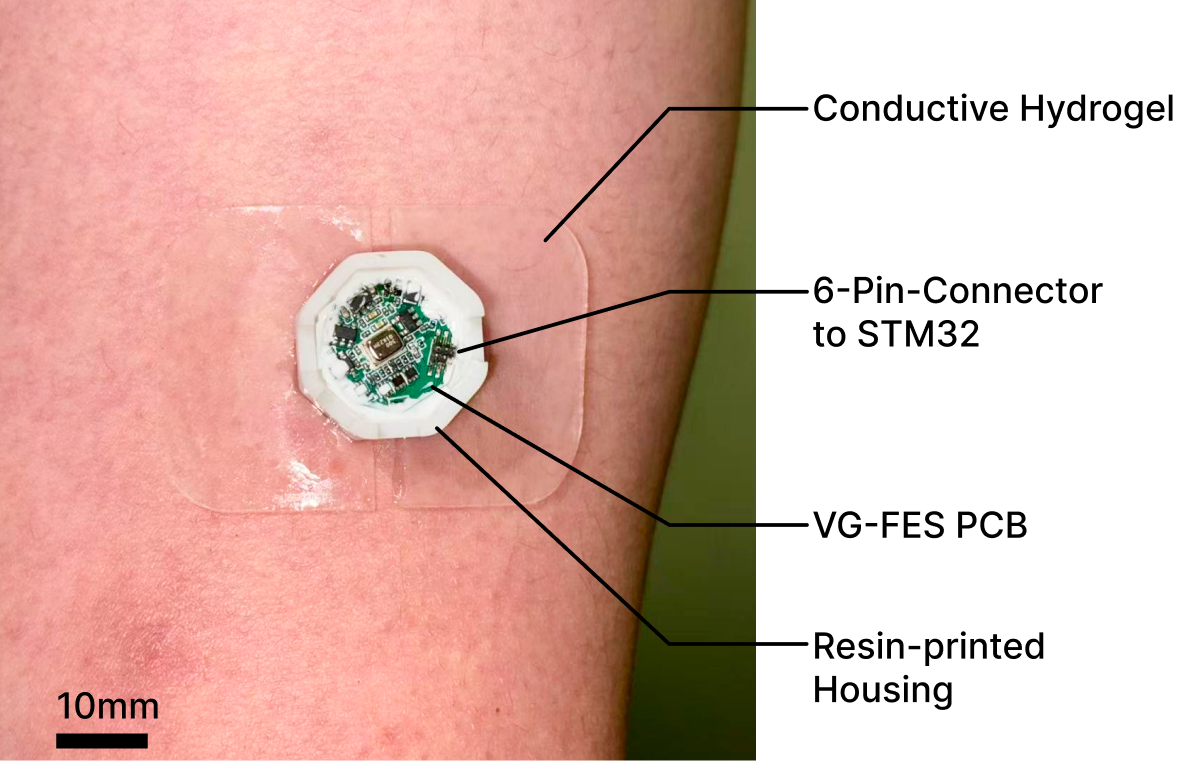}
    \caption{On-skin photograph of the MMG-FES wearable sensor, showing the resin-printed housing, conductive hydrogel electrode, and integrated MMG-FES PCB.Scale bar: 10 mm.}
    \label{s}
\end{figure}

Most wearable FES systems rely on open-loop control or sensing modalities that cannot operate reliably during active stimulation, with EMG in particular being strongly corrupted by stimulation-induced electrical artifacts. In contrast, the MMG–FES module as shown in Figure.~\ref{s} is structurally designed to co-locate MMG sensing and electrical stimulation within a single wearable unit, leveraging the intrinsic immunity of low-frequency mechanical MMG signals to electrical interference to enable continuous, closed-loop muscle-state monitoring during stimulation.

\begin{figure}[ht]
    \centering
    \includegraphics[width=0.9\linewidth]{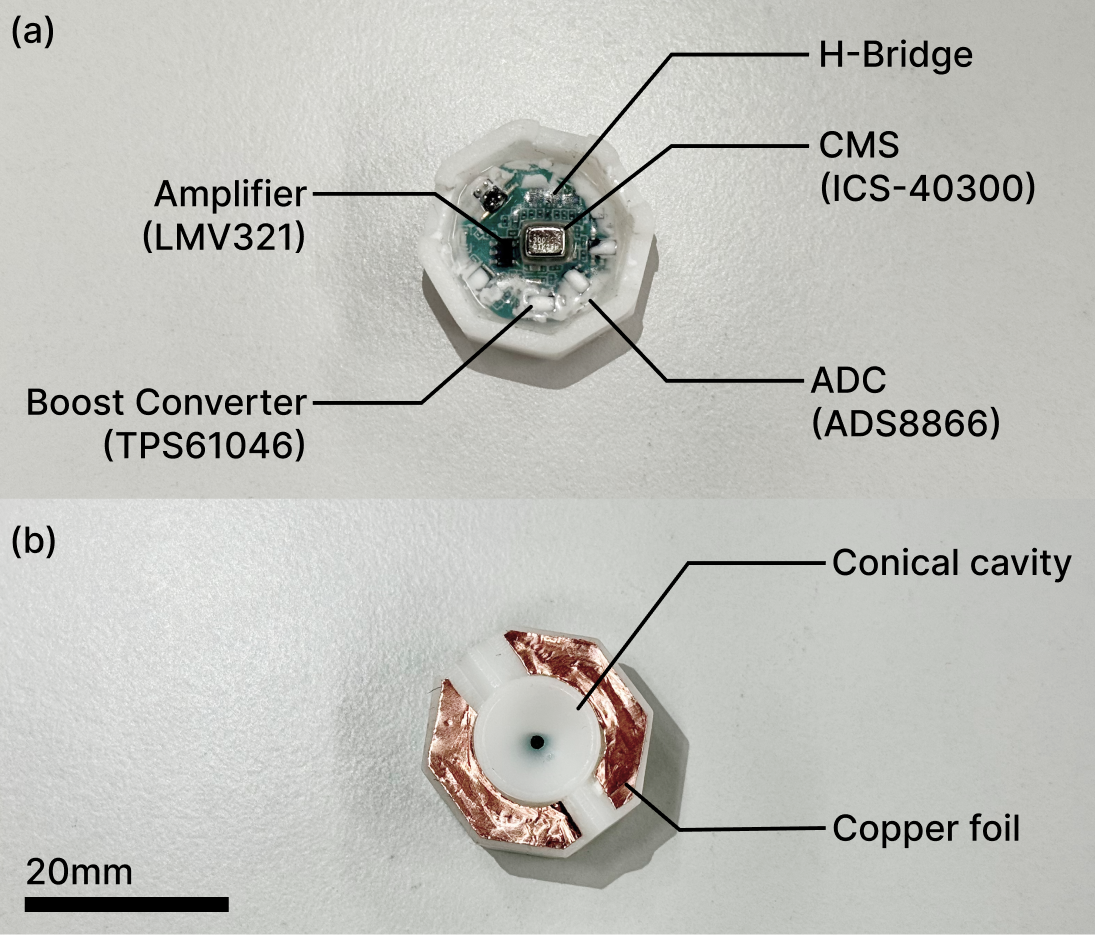}
    \caption{(a) Top view of the MMG-FES PCB showing the CMS microphone (ICS-40300), analog front-end amplifier (LMV321), ADC (ADS8866), boost converter (TPS61046), and H-bridge circuit. (b) Bottom view of the module showing the conical acoustic cavity and copper foil layer. Scale bar: 20 mm.}
    \label{ss}
\end{figure}

The revised MMG–FES system adopts a fully consolidated single-layer PCB architecture, in which the complete MMG sensing chain and functional electrical stimulation (FES) circuitry are co-located on the same board, while wireless communication is deliberately removed and offloaded to an external processing unit as shown in Figure.~\ref{ss} (a). The sensing front end is centered around a CMS capacitive MEMS microphone (ICS-40300, TDK InvenSense), followed by a low-noise analog front-end amplifier (LMV321) that provides initial gain and signal conditioning. The amplified MMG signals are digitized using a high-resolution external ADC (ADS8866) to preserve low-amplitude mechanical vibration details critical for fatigue-related feature extraction. In parallel, the board integrates a complete stimulation subsystem, comprising a buck–boost power stage based on the TPS61046 to provide an adjustable compliance voltage, a programmable constant-current control loop, and a discrete full H-bridge circuit that enables symmetric bipolar FES waveforms. All sensing, stimulation, and power paths are routed with short, symmetric traces to minimize electromagnetic coupling and reduce mutual interference between low-level MMG acquisition and high-voltage stimulation switching. By eliminating on-board Bluetooth and related RF components, the PCB achieves reduced layout complexity, lower electromagnetic noise, and improved signal integrity, while also decreasing power consumption and thermal load. Control and real-time adaptation are handled locally by the STM32F411CE microcontroller, which performs MMG feature extraction and stimulation timing control, whereas high-level computation and machine-learning inference are executed on an external unit via a wired interface. This tightly integrated, communication-agnostic PCB design enhances robustness, simplifies wearable packaging, and provides a stable hardware foundation for closed-loop MMG-driven FES operation.

The MMG-FES integrated sensor adopts a compact hexagonal enclosure designed to balance mechanical robustness, skin conformity, and wearable comfort. The housing has a parallel-edge width of 23 mm, a height of 12 mm, and a total mass of approximately 4 g, minimizing inertial loading on the skin during dynamic movements. The enclosure is fabricated via resin-based additive manufacturing to achieve sufficient structural strength while reducing overall weight and enabling precise control of internal geometry.

As shown in Figure.\ref{ss}, the device employs a two-layer vertical structure. The PCB is bonded to the inner surface of the upper shell using high-strength adhesive, with the CMS microphone inlet precisely aligned to a conical cavity at the bottom of the housing. This conical geometry enhances the coupling efficiency and directionality of low-frequency muscle-induced mechanical vibrations, consistent with prior findings \cite{11276866}. The upper cavity is encapsulated with Dragon Skin silicone to seal the PCB, improving sweat resistance and mechanical stability while preserving MMG sensitivity. A lateral feedthrough routes stimulation leads to the lower electrode without interfering with the sensing pathway. Notably, the apex of the conical structure extends slightly below the plane of the stimulation electrode, ensuring firm skin contact and eliminating air gaps at the sensor–skin interface, which is critical for stable MMG acquisition during movement and stimulation.

Electrical stimulation is delivered through a commercially available conductive hydrogel electrode positioned beneath the sensing module. The hydrogel provides adequate ionic conductivity and intrinsic adhesiveness, allowing the sensor to remain securely attached to the lower limb over extended wearing periods. This stable attachment helps maintain consistent contact pressure and reliable stimulation delivery under dynamic conditions, while supporting robust MMG sensing through sustained mechanical coupling.

The electrical stimulation module delivers biphasic square-wave pulses at three preset frequencies: 5Hz, 15Hz, and 25Hz. These frequencies were selected to represent low-, moderate-, and relatively higher-frequency stimulation conditions for evaluating frequency-dependent muscle mechanical responses. The duty cycle was fixed at 50\%. To ensure safety under transcutaneous stimulation, the charge per pulse was constrained below $7\,\mu\mathrm{C}$, the maximum RMS output amplitude did not exceed 10V, and the single-pulse energy delivered through the skin electrodes was limited to below 300mJ.

\subsection{Frequency-Response Validation of the MMG-FES Sensing Module}

To assess whether the CMS-based sensing module in the MMG-FES system provides sufficient low-frequency sensitivity for mechanomyography (MMG) acquisition, a controlled frequency-response experiment was conducted using a bone-conduction speaker as the mechanical excitation source. The fabricated sensor was mounted on a 3 cm diameter, 8 mm thick Dragon Skin silicone pad to approximate human soft-tissue mechanical impedance. Sinusoidal vibrations were applied over 20–200 Hz at two excitation amplitudes (50 mV and 200 mV), and the sensor outputs were digitized and converted to magnitude values in dBFS relative to the ADC full-scale range, enabling device-level characterization independent of human variability.

\begin{figure}[ht!] 
\centering
\includegraphics[width=3.7in]{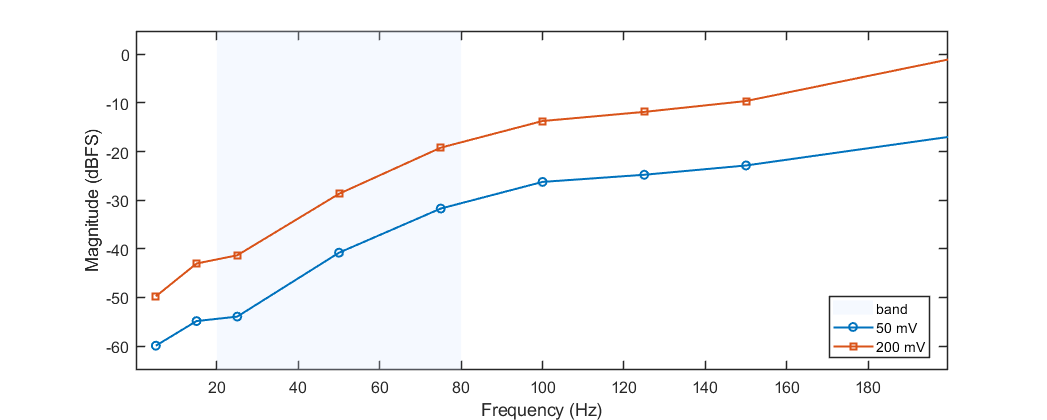}
\caption{
Frequency response of the MMG-FES sensor mounted on a Drogen Skin pad under 50 mV and 200 mV excitation. The shaded 0–200 Hz region denotes the MMG band, where the sensor maintains stable low-frequency sensitivity.
}
\label{fre}
\end{figure}

As shown in Figure.\ref{fre}, the sensor exhibits a clear monotonic increase in response magnitude with frequency for both excitation levels. Under 50 mV excitation, the output rises from approximately -60 dBFS at low frequencies to -18 dBFS at 200 Hz, while the 200 mV excitation produces consistently higher magnitudes, approaching -2 dBFS at the upper end of the tested band. The two curves follow near-parallel trends across the full frequency range, indicating stable amplitude scaling and the absence of strong nonlinear or resonance effects introduced by the enclosure or acoustic coupling structure.

Within the dominant MMG frequency band (20-80 Hz), the CMS sensor demonstrates robust low-frequency detectability, with magnitudes increasing from approximately -55 dBFS to --30 dBFS at 50 mV excitation and from -42 dBFS to -18 dBFS at 200 mV excitation. No resonance peaks or sharp attenuation dips are observed in this range, confirming a smooth and predictable response to muscle-relevant vibrations. These results verify that the CMS module meets the bandwidth and sensitivity requirements for MMG monitoring, providing a reliable device-level foundation for subsequent on-skin experiments and closed-loop MMG-FES control.

\subsection{Performance-Oriented Evaluation of MMG-FES Sensing under Electrical Stimulation}

As an initial validation step, we conducted a baseline comparison between the proposed MMG-FES vibration-based sensing module and a conventional surface electromyography (EMG) system under identical voluntary motion conditions. This comparison aims to establish reference signal characteristics in the absence of electrical stimulation and to assess the temporal and structural differences between the two sensing modalities.

\begin{figure}[ht] 
\centering
\includegraphics[width=0.9\linewidth]{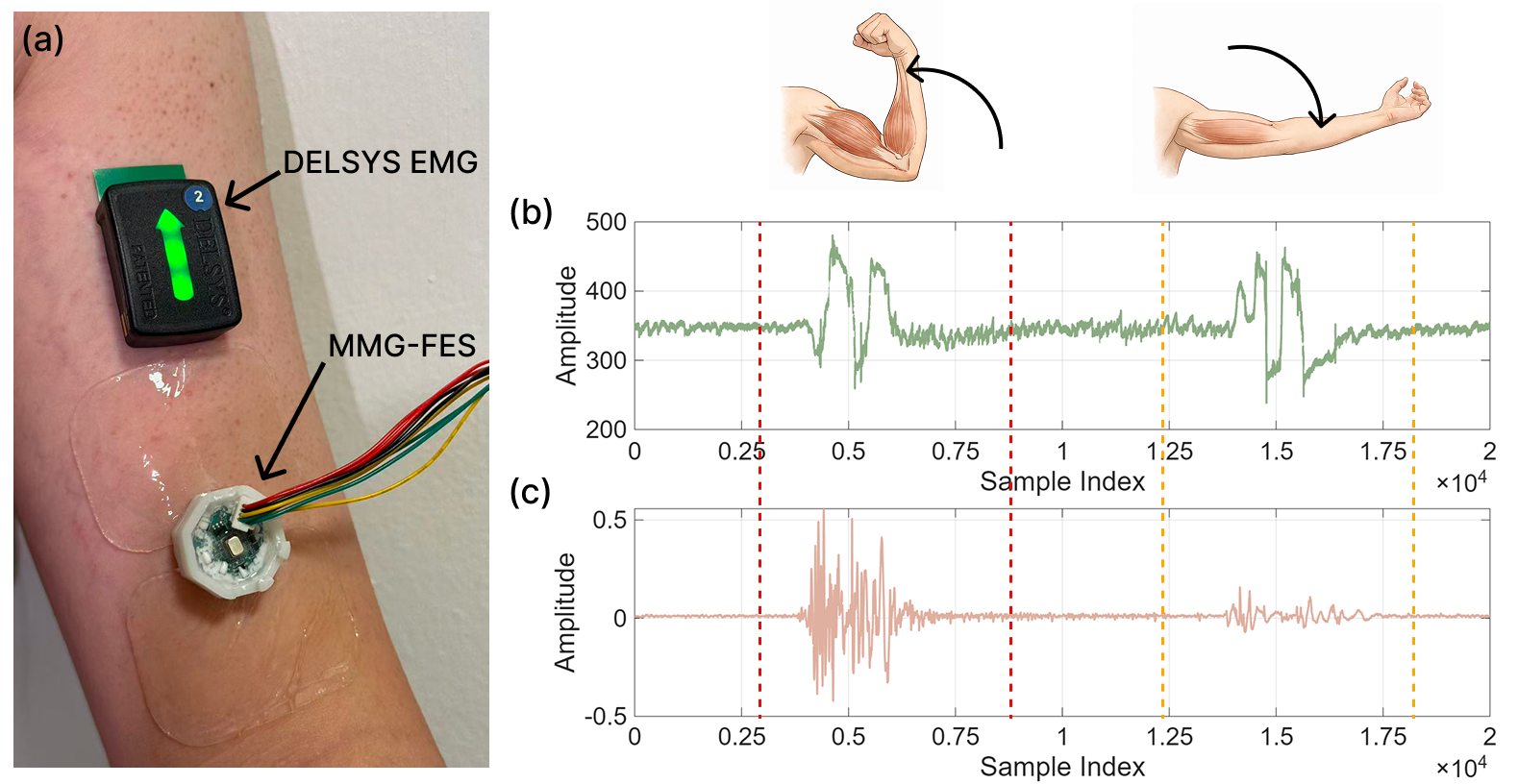}
\caption{
(a) Adjacent placement of the Delsys Trigno Avanti EMG and proposed MMG-FES sensor on the biceps belly. (b) MMG-FES time-domain signal during voluntary elbow flexion (no stimulation). (c) Corresponding EMG time-domain signal under the same condition.
}
\label{normal}
\end{figure}

EMG data were acquired using a Delsys Trigno Avanti Sensor, a high-fidelity wireless surface EMG module widely adopted in biomechanics and neuromuscular research. The Trigno Avanti system captures muscle electrical activity via integrated surface electrodes and transmits digitized signals to the acquisition unit with high signal-to-noise ratio and temporal resolution. The use of this system provides a well-established and reliable EMG baseline for comparison while maintaining consistent experimental conditions with the MMG-FES sensing module, as shown in Figure.\ref{normal} (a). For all baseline experiments, both the MMG-FES sensor and the EMG electrodes were placed over the biceps brachii muscle. The sensing location was centered on the muscle belly, approximately at the midpoint between the acromion and the cubital fossa, ensuring consistent muscle coverage and comparable sensor–muscle coupling across trials.

Both sensing modalities were able to capture voluntary elbow motion consisting of an initial elbow flexion followed by elbow extension, corresponding to the relative movement between the forearm and upper arm. Clear signal variations associated with muscle activity were observed in both recordings under sampling frequency 2150 Hz, indicating that the performed motion was consistently detected under baseline conditions. Despite the presence of measurement noise, distinct motion-related patterns remained identifiable in the acquired signals.

For the EMG signal as shown in Figure.\ref{normal} (b), although a relatively high noise level was present, muscle contraction phases were clearly reflected by rapid increases in signal amplitude. However, the EMG response during muscle relaxation was comparatively weak and less structured, showing limited sensitivity to relaxation-related mechanical changes. In contrast, the MMG-FES sensor exhibited pronounced signal variations during both muscle contraction and relaxation as shown in Figure.\ref{normal} (c). Even in the presence of background noise, MMG-FES measurements showed substantially larger relative changes associated with muscle activity, indicating sufficient sensitivity and signal-to-noise separation to capture dynamic muscle motion.

\begin{figure}[!ht] 
\centering
\includegraphics[width=0.9\linewidth]{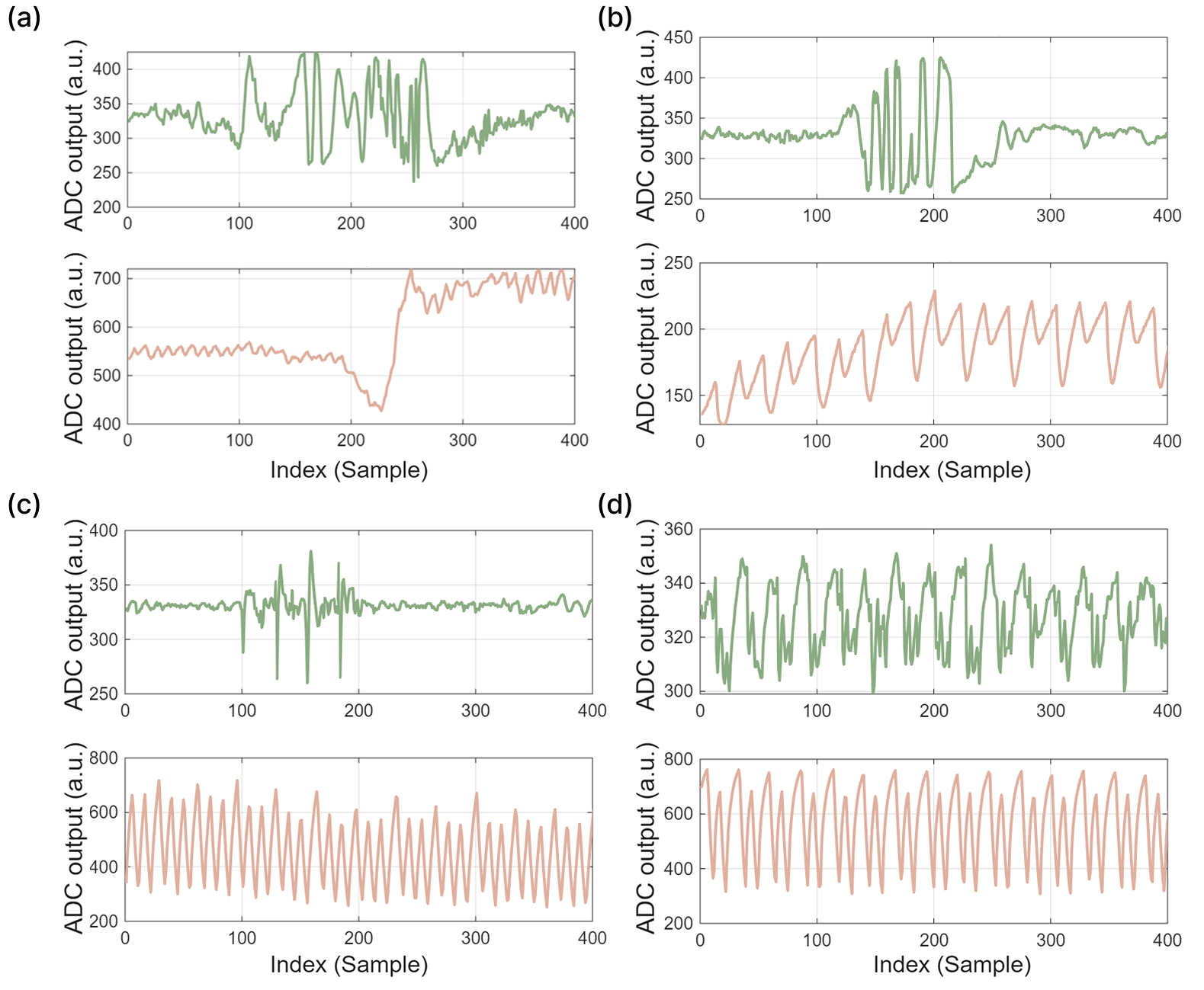}
\caption{
MMG-FES (top) and EMG (bottom) signals during standing under FES. (a)–(c) 5, 15, and 25 Hz; (d) stronger stimulation. EMG is dominated by stimulation-locked oscillations, whereas MMG-FES preserves movement-related vibration patterns.
}
\label{FES}
\end{figure}

Following the baseline comparison under non-stimulated conditions, we evaluated the behavior of the MMG-FES sensor and a conventional EMG device in the presence of functional electrical stimulation (FES). Electrical stimulation was applied at three representative frequencies—5 Hz, 15 Hz, and 25 Hz—while the same voluntary lower-limb movement was performed. Representative signal segments are shown Figure.\ref{FES}. For this baseline experiments, both the MMG-FES sensor and the EMG electrodes were placed over the quadriceps muscle group, with the sensing location centered on the muscle belly approximately 15 cm above the superior border of the patella, ensuring consistent muscle coverage and comparable sensor–muscle coupling across trials.

During a voluntary standing-up movement (sit-to-stand transition), we evaluated the behavior of the MMG-FES sensor and a conventional EMG device under functional electrical stimulation at different frequencies. At 5 Hz stimulation (Figure.\ref{FES}(a)), the MMG-FES vibration signal remains sensitive to muscle activity and exhibits clear movement-related fluctuations. In contrast, the EMG signal already shows pronounced periodic oscillations synchronized with the stimulation pulses, with movement-related changes partially visible but strongly contaminated by stimulation-induced artifacts. As the stimulation frequency increases to 15 Hz (Figure.\ref{FES}(b)), the MMG-FES signal continues to preserve a clean temporal structure with distinguishable movement-related features, whereas the EMG signal becomes dominated by periodic oscillations corresponding to the stimulation frequency, leaving only a slow baseline drift associated with the standing-up movement. At 25 Hz stimulation (Figure.\ref{FES}(c)), the MMG-FES signal still exhibits interpretable vibration patterns associated with muscle activity, while the EMG signal is almost entirely governed by high-amplitude stimulation-induced oscillations, rendering voluntary movement recognition effectively infeasible.These results in Fig.~\ref{FES}(a–c) primarily demonstrate the limitations of conventional EMG sensing under electrical stimulation and motivate the need for an alternative sensing modality capable of operating in electrically active environments.

Crucially, beyond tolerating electrical stimulation, the MMG-FES sensor enables direct observation of stimulation-induced muscle vibration in the absence of voluntary movement, which constitutes a mechanically meaningful physiological response rather than an electrical artifact. This capability is highlighted in Figure.\ref{FES}(d), where increasing stimulation intensity produces structured, repeatable vibration patterns in the MMG-FES signal that reflect the intrinsic mechanical response of the muscle tissue to electrical excitation, without any intentional motor task being performed. While the EMG signal recorded under the same condition is dominated by large-amplitude, stimulation-synchronized oscillations and frequently approaches the saturation limits of the analog-to-digital converter, the MMG-FES signal remains within a stable dynamic range and preserves interpretable temporal features.

These observations indicate that MMG-FES sensing is not merely robust to electrical stimulation, but is specifically suited for capturing the mechanical manifestations of electrically induced muscle activation, providing a direct window into muscle vibration dynamics that are inaccessible to conventional EMG-based measurements.

\section{Experimental Validation of MMG-FES Sensing and Therapeutic Functionality}
\label{Data}

\subsection{Muscle Selection and Placement Strategy}

To demonstrate the practical relevance and broad applicability of the proposed MMG-FES system, nine representative skeletal muscles were selected based on their frequent use in electrical stimulation and vibration-based therapeutic interventions. The selection was guided not by anatomical specificity, but by therapeutic significance, surface accessibility, and their established roles in functional rehabilitation.

\begin{figure}[!htbp]
    \centering
    \includegraphics[width=0.9\linewidth]{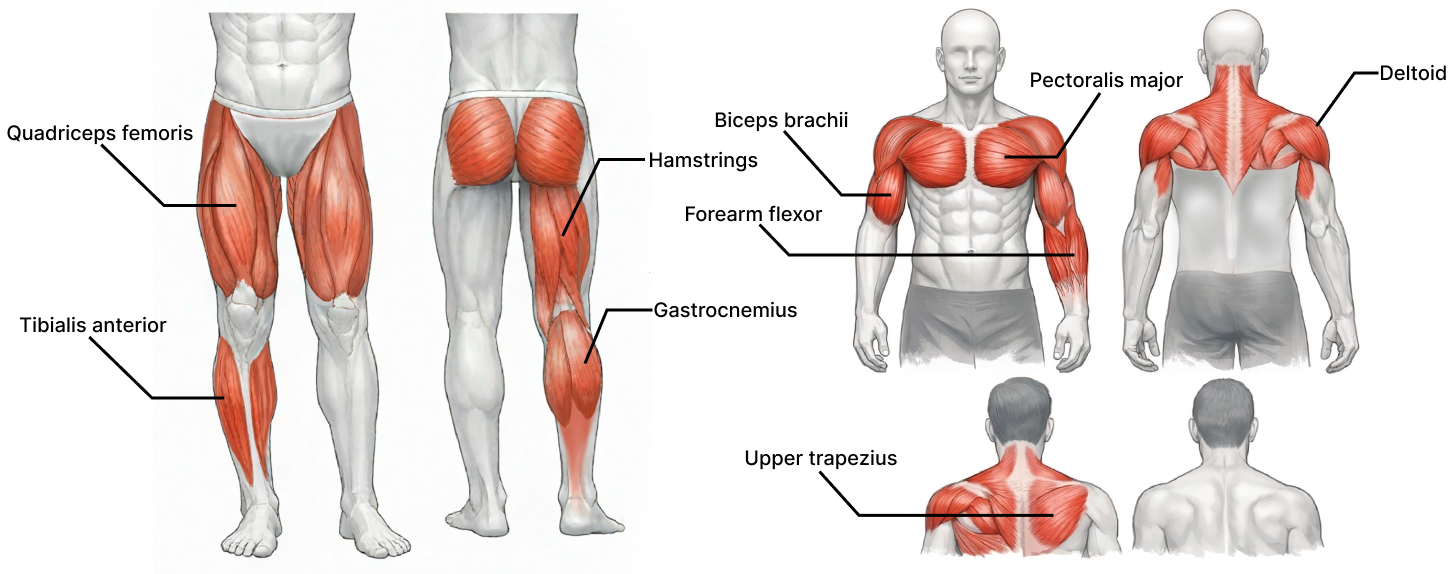}
    \caption{Left: Representative lower-limb muscles below the waist: quadriceps femoris, hamstrings,tibialis anterior and gastrocnemius. Right: Representative upper-limb and shoulder-girdle muscles above the waist: biceps brachii, deltoid, forearm flexor muscle group, pectoralis major and upper trapezius.}
    \label{place}
\end{figure}

Specifically, four lower-limb muscles—quadriceps femoris, hamstrings, tibialis anterior and  gastrocnemius as shown in Figure.\ref{place} (a)—were selected because they are particularly susceptible to disuse atrophy, fatigue, and impaired neural drive following musculoskeletal injury, neurological disorders, or prolonged immobilization, and therefore constitute primary targets in electrical stimulation–based rehabilitation aimed at restoring functional activation and preventing degeneration \cite{Flodin2022Effects}. In clinical practice, electrical stimulation is commonly applied to these muscles to artificially evoke controlled contractions accompanied by mechanical vibration responses, supporting gait assistance, postural transitions such as sit-to-stand, and correction of foot drop\cite{Li2025Effects}. In these scenarios, maintaining effective activation while avoiding excessive loading requires timely and adaptive adjustment of stimulation parameters.

Similarly, three upper-limb muscles—biceps brachii, deltoid, forearm flexor muscle group as shown in Figure.\ref{place} (b)—were included because they are prone to weakness, spasticity, and rapid fatigue in conditions such as stroke, spinal cord injury, and postoperative recovery\cite{Dolbow2024The}. Electrical stimulation and vibration-based interventions are widely employed to artificially excite these muscles in order to facilitate elbow flexion–extension, restore grasping ability, and support functional manipulation\cite{Schuhfried2012Non-invasive}. Owing to their frequent involvement in upper-limb rehabilitation and their sensitivity to fatigue and overstimulation, these muscles represent clinically meaningful targets for vibration-aware, closed-loop stimulation control.

In addition, two upper-limb and shoulder-girdle muscles—pectoralis majorand upper trapezius as shown in Figure.\ref{place} (b)—were selected due to their central involvement in upper-body force generation, shoulder girdle stabilization, and trunk–upper-limb coordination, as well as their susceptibility to fatigue, overload, and impaired activation in both daily activities and rehabilitation settings\cite{Barnamehei2022Effect}. These large, surface-accessible muscles are frequently targeted in electrical stimulation and vibration-based therapies to artificially excite muscle activity and restore functional pushing, pulling, and postural support\cite{Muyor2022Comparison}. Electrical stimulation of these muscles reliably elicits mechanically observable vibration responses, which are particularly informative for assessing muscle loading and activation state during therapy. Importantly, the proposed MMG-FES system captures local mechanical vibration responses generated by muscle contraction rather than relying on muscle-specific electrophysiological pathways. As a result, system performance does not depend on precise muscle identification or motor-point alignment, enabling flexible deployment across these clinically relevant muscle groups. This muscle-agnostic characteristic supports adaptive, closed-loop stimulation adjustment in a wide range of real-world therapeutic scenarios.

\subsection{Validation of MMG-FES Sensing in Electrically Stimulated Muscles}

To verify that MMG-FES signals recorded under electrical stimulation genuinely reflect muscle-induced mechanical vibration rather than stimulation-related artifacts, we first conducted an auxiliary validation using an inertial measurement unit (IMU) as an independent mechanical sensing reference.

\begin{figure}[!htbp]
    \centering
    \includegraphics[width=0.9\linewidth]{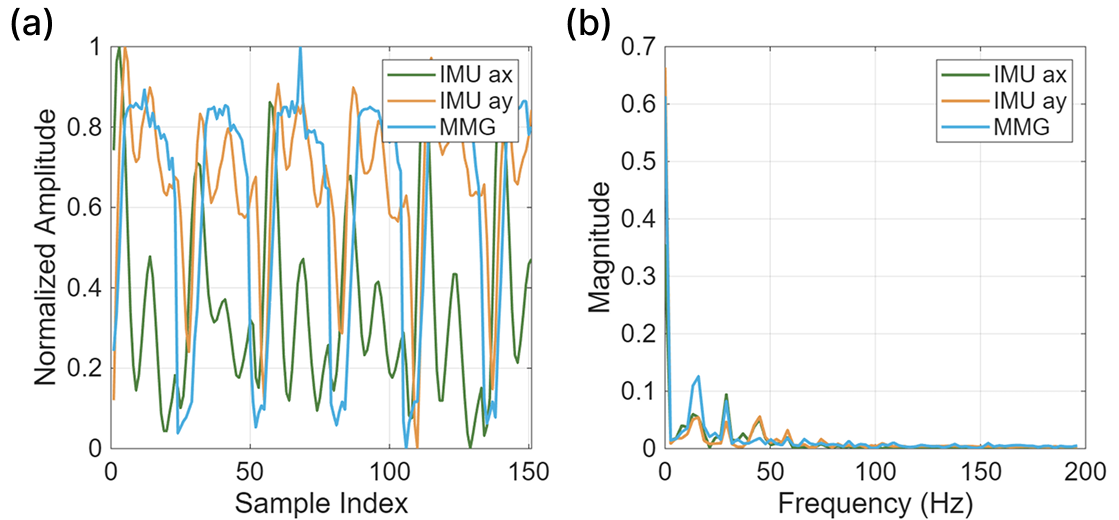}
    \caption{IMU-assisted validation of MMG-FES sensing under electrical stimulation. (a) Normalized time-domain signals from IMU accelerometer axes and MMG-FES during medium-intensity stimulation of the thigh. (b) Corresponding frequency spectra obtained via FFT.}
    \label{IMU}
\end{figure}
As shown in Figure.\ref{IMU}, during medium-intensity electrical stimulation (15 Hz) applied to the biceps brachii, MMG-FES signals exhibited strong temporal and spectral consistency with IMU-measured mechanical vibration. In the time domain, the normalized MMG-FES waveform closely followed the vibration patterns captured by the IMU accelerometer axes, indicating that both modalities responded to the same underlying mechanical events induced by muscle contraction. At the same time, MMG-FES signals contained additional fine-grained vibration components that were less pronounced in the IMU measurements. Frequency-domain analysis further supported this observation, as the dominant frequency components in the MMG-FES spectrum closely matched those extracted from the IMU signals after FFT transformation, particularly within the low-frequency band associated with muscle mechanical activity. Taken together, the agreement observed in both temporal structure and spectral content demonstrates that MMG-FES directly captures muscle-induced mechanical vibration under electrical stimulation, rather than stimulation-related electrical or electromagnetic artifacts.

A total of five able-bodied participants (three males and two females) were recruited for this study. Participants ranged in age from 21 to 25 years, with body mass between 66–85 kg and height between 157–176 cm, covering a representative range of body compositions and muscularity levels (Table~\ref{tab:participants}). All participants reported no history of neuromuscular or musculoskeletal disorders, mobility impairments, or skin conditions that could interfere with surface electrical stimulation or vibration sensing. Prior to participation, all individuals were informed of the experimental procedures and provided written informed consent.

\begin{table}[!htbp]
\centering
\caption{Physical characteristics of participants involved in the cross-subject consistency analysis}
\label{tab:participants}
\begin{tabular}{lcccc}
\hline
\textbf{Participant} & \textbf{Sex} & \textbf{Weight(kg)} & \textbf{Height(cm)} & \textbf{Body composition} \\
\hline
P1 & M & 76 & 175 & Average \\
P2 & M & 80 & 168 & Higher adiposity \\
P3 & M & 85 & 176 & Solid \\
P4 & F & 73 & 167 & Solid \\
P5 & F & 66 & 157 & Lean \\
\hline
\end{tabular}
\end{table}

Data were collected under two non-fatigued muscle states. In the Rest condition, participants remained fully relaxed without voluntary contraction. In the IsoLow condition, they maintained a weak, constant isometric contraction (~10\% perceived maximal effort), monitored verbally and visually. All trials were performed in a seated, upright position to ensure postural stability, and participants avoided voluntary movements so that signals primarily reflected stimulation-induced mechanical responses. Functional electrical stimulation (FES) was delivered using three intensity levels (Low, Mid, High) across all muscles and participants. Adequate rest was provided between trials to minimize fatigue and maintain repeatability.

\begin{figure*}[!htbp] 
\centering
\includegraphics[width=1.0\textwidth]{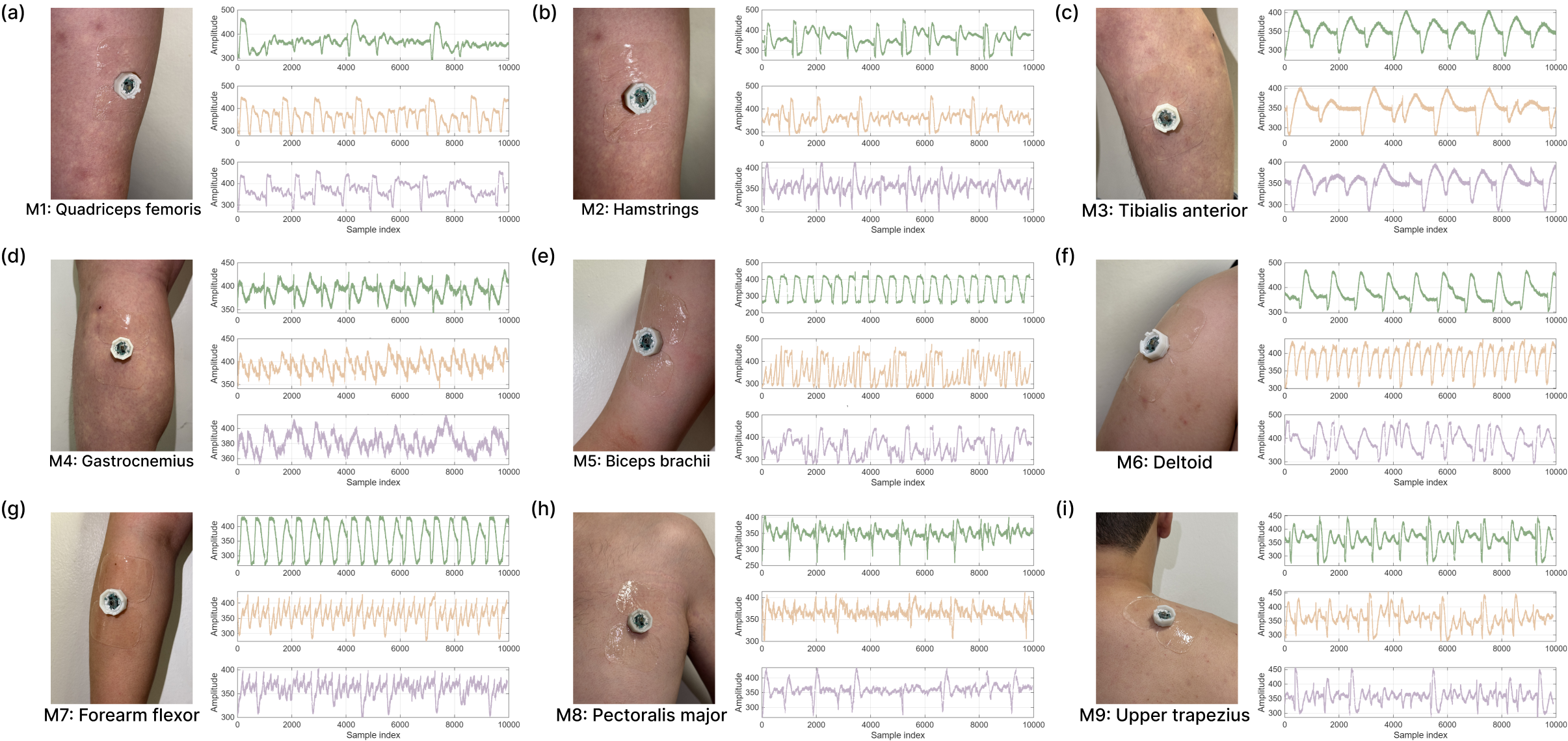}
\caption{
     MMG-FES signals recorded from nine muscle groups under three stimulation levels. (a) Quadriceps femoris, (b) Hamstrings, (c) Tibialis anterior, (d) Gastrocnemius, (e) Biceps brachii, (f) Deltoid, (g) Forearm flexor, (h) Pectoralis major, (i) Upper trapezius. For each muscle, MMG-FES signals are shown at low, mid, and high stimulation levels.
}
\label{data}
\end{figure*}

Across all nine muscle groups, the MMG-FES sensor captured clear and structured vibration signals under electrical stimulation (Fig.~\ref{data}), demonstrating high sensitivity to mechanically relevant muscle activity. Signal amplitudes closely matched participants’ perceived contraction strength, indicating that the recordings reflected genuine mechanical responses rather than electrical artifacts. Distinct patterns were observed across muscles, consistent with muscle-specific mechanical properties and sensor–tissue coupling, yet responses remained stable and repeatable across anatomical locations. Stimulation intensity systematically modulated the signals: low intensity produced intermittent pulses, medium intensity generated denser vibrations, and high intensity introduced larger amplitudes with increased high-frequency components, reflecting sustained excitation.

Overall, MMG-FES reliably captures muscle vibration, preserves muscle-specific signatures, and sensitively reflects stimulation intensity, supporting its versatility for electrically stimulated neuromuscular systems.

\begin{figure}[!htbp]
    \centering
    \includegraphics[width=1.0\linewidth]{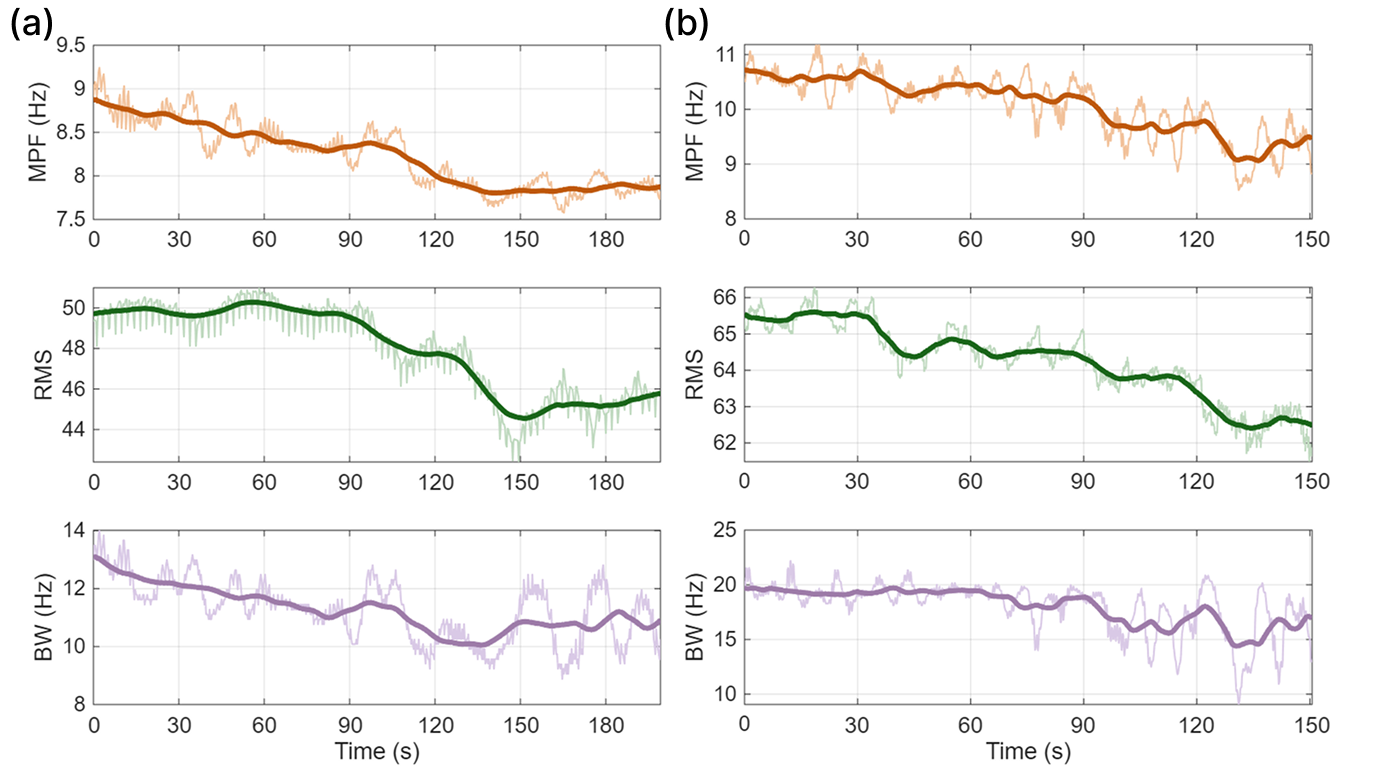}
    \caption{Temporal evolution of MMG features during electrical stimulation. (a) Quadriceps femoris and (b) biceps brachii. MPF, RMS, and spectral bandwidth (BW) decrease during early stimulation, reflecting progressive muscle relaxation.}
    \label{mpf}
\end{figure}

\subsection{MMG-Based Feature Analysis of Neuromuscular Response Under FES}
To further characterize the neuromuscular response during functional electrical stimulation (FES), three complementary descriptors were extracted from the MMG signal: the mean power frequency (MPF), the root mean square amplitude (RMS), and the spectral bandwidth (BW). These features respectively quantify the dominant vibration frequency, mechanical vibration amplitude, and spectral energy distribution of stimulation-evoked muscle activity.

The mean power frequency (MPF) of the MMG signal was computed as a frequency-domain descriptor of stimulation-evoked mechanical activity. MPF is defined as

\begin{equation}
\label{eq:mpf}
\mathrm{MPF}=\frac{\sum\limits_{f=f_l}^{f_h} f\,P(f)}{\sum\limits_{f=f_l}^{f_h} P(f)},
\end{equation}

where $P(f)$ denotes the power spectral density (PSD) of the MMG signal at frequency $f$, and $[f_l,f_h]$ is the selected analysis band. As the power-spectrum centroid, MPF quantifies shifts in the dominant vibration frequency associated with changes in motor-unit mechanical activity.

The root mean square amplitude (RMS) was used to measure the overall magnitude of muscle mechanical vibration and is calculated as

\begin{equation}
\label{eq:rms}
\mathrm{RMS}=\sqrt{\frac{1}{N}\sum_{n=1}^{N} x_n^2},
\end{equation}

where $x_n$ represents the $n$th MMG sample within a time window of length $N$. RMS reflects the intensity of mechanical oscillations generated by electrically evoked muscle contractions.

To characterize the spectral distribution of vibration energy, the spectral bandwidth (BW) was computed as the difference between the 95\% and 5\% cumulative spectral energy frequencies:

\begin{equation}
\label{eq:bw}
\mathrm{BW}=f_{95}-f_{5},
\end{equation}

where $f_{5}$ and $f_{95}$ denote the frequencies at which the cumulative power spectrum reaches 5\% and 95\% of the total spectral energy, respectively. BW therefore represents the spread of vibration energy across the analyzed frequency range.

As shown in Fig.~\ref{mpf}, distinct temporal evolution patterns of these MMG descriptors were observed in both the quadriceps femoris and biceps brachii under 5~Hz and 25~Hz stimulation. During the early stimulation phase (Fig.~\ref{mpf}a--b), MPF, RMS, and BW exhibited a gradual decrease. This behavior reflects progressive muscle relaxation induced by electrical stimulation. As muscle tension decreases, the effective stiffness of the muscle–tendon unit is reduced. From a mechanical vibration perspective, the dominant vibration frequency is approximately related to system stiffness by

\begin{equation}
f \propto \sqrt{\frac{k}{m}},
\end{equation}

where $k$ represents the effective stiffness and $m$ the equivalent mass of the vibrating system. A reduction in stiffness therefore shifts vibration energy toward lower frequencies, resulting in a decrease in MPF. Simultaneously, the reduction in mechanical tension leads to lower vibration amplitudes (RMS decrease) and a narrowing of spectral bandwidth, indicating fewer high-frequency vibration components. Participants reported this stage as increasingly relaxed and comfortable.

\begin{figure}[t]
    \centering
    \includegraphics[width=0.9\linewidth]{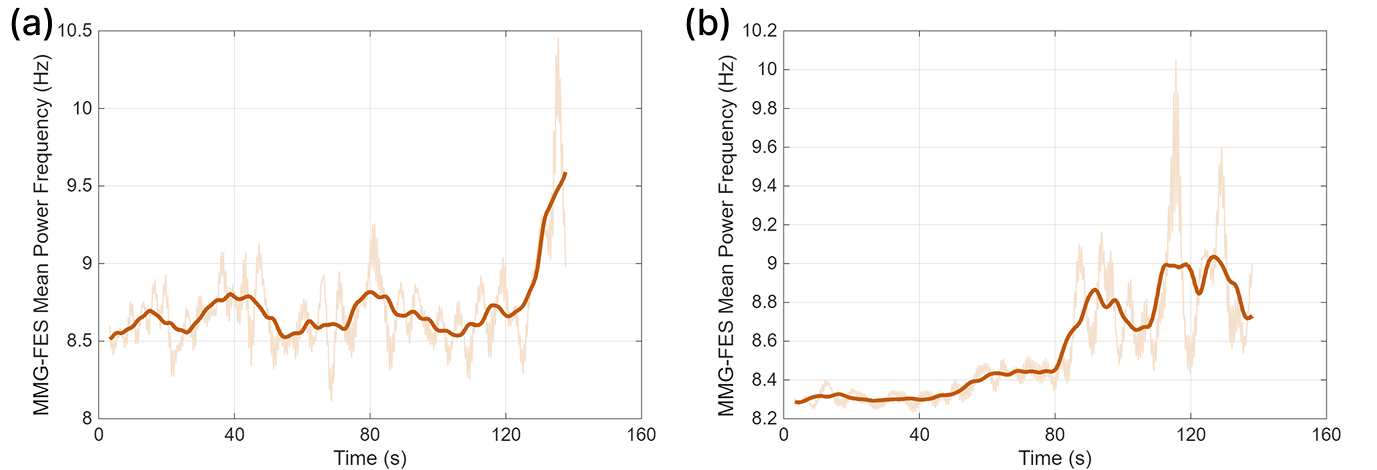}
    \caption{Temporal evolution of MMG mean power frequency (MPF) during electrical stimulation. (a) Quadriceps femoris and (b) biceps brachii. Prolonged stimulation leads to a gradual increase in MPF, indicating fatigue-related changes in muscle mechanical stiffness.}
    \label{mpf}
\end{figure}

However, with prolonged stimulation (Fig.~\ref{mpf}c--d), a clear increase in MPF was observed. This transition corresponds to the onset of stimulation-induced neuromuscular fatigue. Fatigue-related metabolic accumulation and reduced cross-bridge efficiency increase passive tension and modify the viscoelastic properties of the muscle, thereby elevating muscle stiffness and internal damping \cite{10.1007/978-981-16-0866-7_103}. The increase in stiffness shifts the vibration spectrum toward higher frequencies, producing a systematic rise in MPF. Participants simultaneously reported sensations of soreness or ``acidic/numb'' discomfort, consistent with fatigue accumulation.

Comparison between pre- and post-exercise sessions revealed a significant elevation in fatigue-related MPF indices following sustained mechanical loading, confirming that exercise-induced stiffness alters vibration dynamics. Collectively, these findings demonstrate that MMG-derived descriptors capture both relaxation-dominant responses and fatigue-related mechanical transitions.

Therefore, real-time monitoring of MMG spectral features may serve as a quantitative feedback variable for adaptive regulation of stimulation duration and intensity, enabling fatigue-aware and patient-specific FES control.

\section{Experimental Validation in Real-World Conditions}

\subsection{Fatigue Features Capture using EMG and MMG-FES Signals}

To validate the fatigue recognition capability of the proposed MMG-FES sensing framework, five healthy participants were recruited to perform 80 consecutive squat actions within a predefined time window. During the task, surface electromyography (EMG) signals and vibration signals acquired by the MMG-FES sensor were synchronously recorded. From the complete set of 80 actions, 20 trials were randomly selected while preserving their temporal order for subsequent fatigue analysis. EMG-based median frequency (MF) and mean power frequency (MPF) were employed as established reference indicators for muscle fatigue during voluntary contractions.

\begin{figure}[!htbp]
    \centering
    \includegraphics[width=0.9\linewidth]{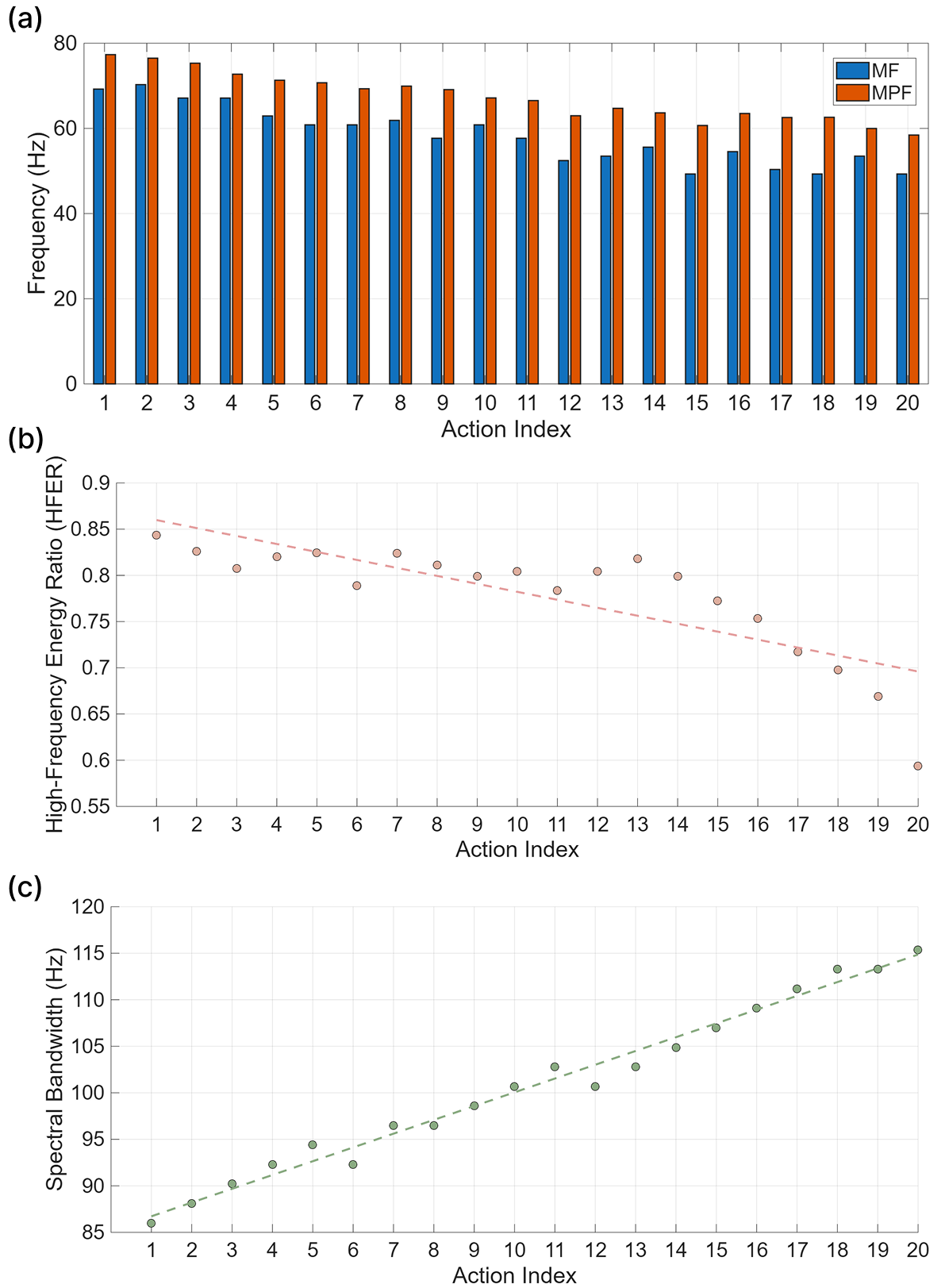}
    \caption{Fatigue-related spectral features extracted from EMG and MMG-FES signals during repeated squat actions. (a) EMG median frequency (MF) and mean power frequency (MPF) decrease progressively with action index, indicating muscle fatigue. (b) High-frequency energy ratio (HFER) of MMG-FES vibration signals shows a decreasing trend across actions. (c) MMG-FES spectral bandwidth (BW) increases progressively, reflecting fatigue-related changes in vibration frequency content.}
    \label{fig:fatigue}
\end{figure}

As shown in Fig.~\ref{fig:fatigue}(a), both MF and MPF exhibited a progressive decrease across action indices. This trend is consistent with well-established EMG fatigue characteristics reported in the literature and reflects a spectral shift toward lower frequencies under sustained muscle activation, which is commonly associated with motor unit conduction velocity reduction and fatigue-induced neuromuscular changes.

In parallel, features were extracted from the MMG-FES vibration signals to characterize fatigue-related mechanical responses. The high-frequency energy ratio (HFER) was defined as
\begin{equation}
\mathrm{HFER} = \frac{\sum_{f \ge f_c} P(f)}{\sum_{f=f_{\min}}^{f_{\max}} P(f)},
\end{equation}
where $P(f)$ denotes the power spectral density of the vibration signal and $f_c$ is the predefined high-frequency threshold. As illustrated in Fig.~\ref{fig:fatigue}(b), HFER showed a clear decreasing trend across actions, indicating a gradual reduction in the proportion of high-frequency vibration energy as fatigue accumulated.

In contrast, the spectral bandwidth (BW) of the MMG-FES signal exhibited a monotonic increase with action index, as shown in Fig.~\ref{fig:fatigue}(c). Taken together, the concurrent decrease in high-frequency energy ratio (HFER) and increase in spectral bandwidth (BW) provide complementary mechanical signatures of muscle fatigue captured by the MMG-FES sensor. Rather than reflecting a simple attenuation or amplification of vibration signals, these concurrent trends indicate a redistribution of mechanical vibration energy across the frequency spectrum. Such redistribution is likely associated with fatigue-related changes in muscle--tissue stiffness and damping, which alter the modal structure and energy transmission characteristics of muscle vibrations.

Importantly, these fatigue-dependent spectral features enable real-time fatigue state recognition during movement and form the basis for adaptive electrical stimulation strategies, allowing stimulation parameters to be dynamically adjusted according to the detected fatigue level. These MMG-derived trends are particularly meaningful because they are observed alongside the well-established decrease in EMG-derived MF and MPF during repeated contractions, providing a physiologically grounded reference for fatigue characterization. The consistent co-evolution between EMG and MMG features suggests that the observed MMG spectral changes are not merely task-dependent signal variations, but are closely associated with underlying fatigue-induced neuromuscular alterations. Therefore, MMG spectral features can serve as reliable indicators for fatigue state recognition, particularly in scenarios where EMG becomes unreliable due to stimulation artifacts during active FES.

\subsection{Fatigue-State Classification Based on MMG-FES Signals}

To investigate whether electrically evoked MMG signals can reliably encode neuromuscular fatigue information during stimulation, a supervised learning framework was developed using a cross-condition data acquisition strategy. In this approach, fatigue states were first identified from voluntary movement signals and then recognized from MMG signals recorded during electrical stimulation. This strategy separates fatigue labeling from stimulation sensing and reduces potential circular reasoning.

\begin{table}[!htbp]
\centering
\caption{Cross-condition fatigue labeling and learning strategy}
\label{tab:fatigue_labeling}
\begin{tabular}{llll}
\hline
Domain & Condition & Features & Role \\
\hline
Movement & Voluntary motion & BW, HFER & Labeling \\
Stimulation & Electrical stimulation & MPF, RMS, BW & Classification \\
\hline
\end{tabular}
\end{table}

As summarized in Table~\ref{tab:fatigue_labeling}, fatigue states were first characterized using vibration descriptors derived from voluntary motion. Based on this characterization, fatigue labels during electrical stimulation were assigned according to stimulation duration, corresponding to early, intermediate, and late phases of the FES protocol.

Three representative muscle groups were selected for evaluation: quadriceps femoris (M1), biceps brachii (M5), and pectoralis major (M8). For each muscle, three fatigue states (\textit{fresh}, \textit{transitional}, and \textit{fatigued}) were defined based on stimulation duration, corresponding to early, intermediate, and late phases of the FES protocol, resulting in nine muscle–fatigue classes. 

Rather than training separate models for each muscle, all data were pooled to perform cross-muscle classification. This design aims to evaluate whether MMG features can capture muscle-independent fatigue characteristics, thereby assessing the robustness and generalizability of the proposed sensing framework across different anatomical locations. A separate set of experiments was conducted under FES conditions with five healthy participants, and data from all subjects were pooled for analysis.

MMG signals recorded during electrical stimulation were first preprocessed using median-based spike removal to suppress impulsive artifacts. The signals were then segmented using a sliding window (3000 samples with a 1000-sample step size). From each valid window, a compact feature vector

\[
\mathbf{x}=[RMS,MPF,BW]
\]

was extracted to characterize vibration amplitude, dominant vibration frequency, and spectral distribution of stimulation-induced muscle activity.

To evaluate the robustness of fatigue recognition, four classical machine learning classifiers were investigated, including logistic regression (LR), support vector machine (SVM), decision tree (DT), and random forest (RF). These algorithms represent different levels of model complexity and nonlinear representation capability, allowing systematic comparison of classification performance for MMG-derived fatigue features.

The LR classifier estimates the probability of each class using a Softmax formulation

\begin{equation}
P(y = k \mid \mathbf{x}) =
\frac{e^{\mathbf{w}_k^\top \mathbf{x} + b_k}}
{\sum_{j} e^{\mathbf{w}_j^\top \mathbf{x} + b_j}},
\end{equation}

where \( \mathbf{w}_k \) and \( b_k \) denote the weight vector and bias of class \(k\). Features were standardized using z-score normalization, and the dataset was divided into training and testing subsets using a stratified 70/30 split.

\begin{table}[!htbp]
\centering
\caption{Performance comparison of machine learning classifiers for fatigue-state recognition}
\label{tab:classifier_accuracy}
\begin{tabular}{lc}
\hline
Classifier & Accuracy \\
\hline
Logistic Regression (LR) & 0.93 \\
Support Vector Machine (SVM) & 0.91 \\
Decision Tree (DT) & 0.91 \\
Random Forest (RF) & \textbf{0.96} \\
\hline
\end{tabular}
\end{table}

As summarized in Table~\ref{tab:classifier_accuracy}, the random forest classifier achieved the highest classification accuracy among the evaluated models, followed by logistic regression, while SVM and decision tree produced slightly lower but comparable results. These findings indicate that ensemble learning methods are particularly effective in capturing nonlinear relationships within MMG-derived fatigue features.

\begin{figure}[!htbp]
\centering
\includegraphics[width=0.9\linewidth]{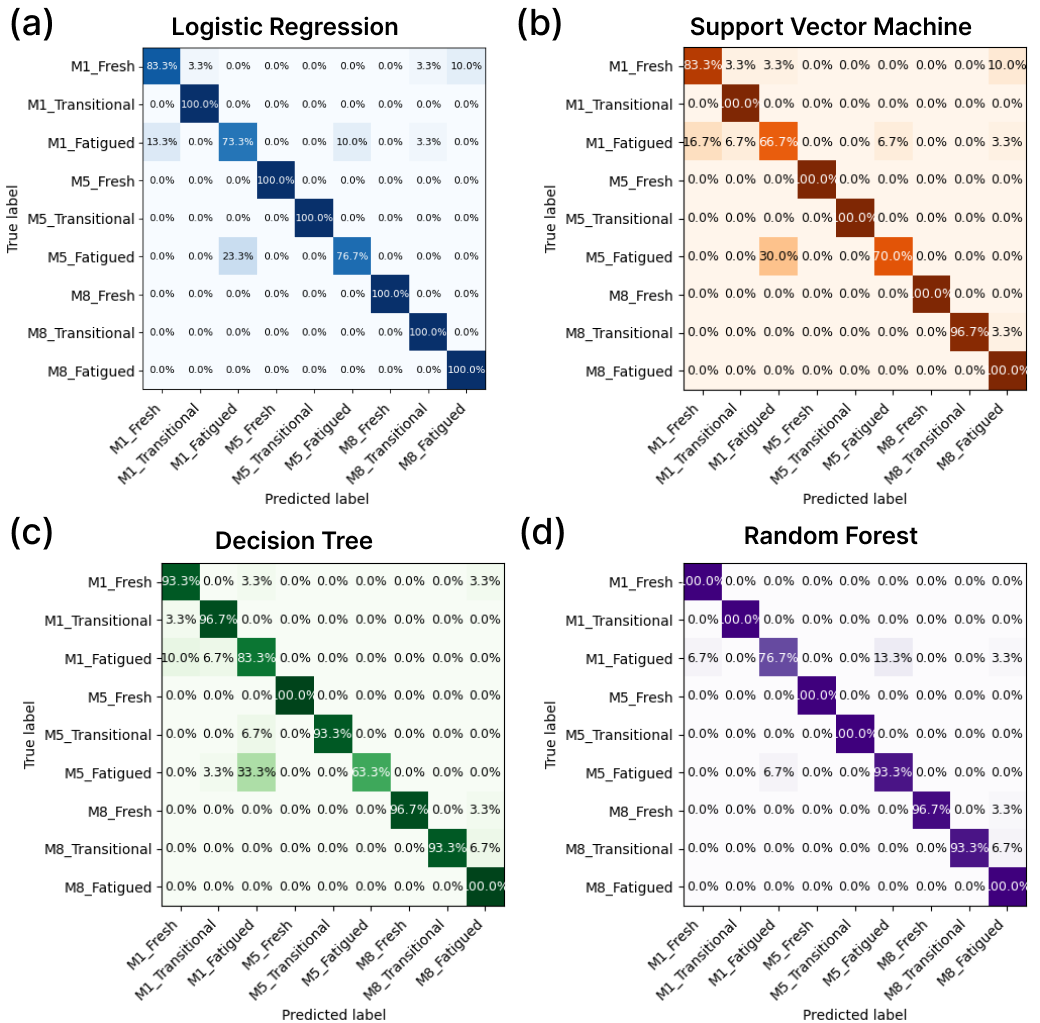}
\caption{Row-normalized confusion matrices for fatigue-state classification using four machine learning algorithms: (a) Logistic Regression, (b) Support Vector Machine, (c) Decision Tree, and (d) Random Forest.}
\label{fig:ML_confusion}
\end{figure}

The confusion matrices shown in Fig.~\ref{fig:ML_confusion} exhibit strong diagonal dominance across all classifiers, indicating reliable discrimination among the nine muscle–fatigue classes. Most classification errors occurred between neighboring fatigue states, particularly between the transitional and fatigued conditions. This behavior reflects the gradual physiological progression of muscular fatigue rather than abrupt state transitions.

Importantly, the fatigue-state classifier also provides a practical basis for stimulation regulation. During voluntary movement, higher BW and HFER values typically indicate increasing muscular load and approaching fatigue. When electrical stimulation is applied, the corresponding MMG descriptors (MPF, RMS, and BW) reflect the mechanical response of the muscle to stimulation. Elevated MPF and RMS values are generally associated with increased vibration intensity and muscle stiffness during fatigue accumulation.

By combining movement-based fatigue calibration with stimulation-derived MMG features, the system can infer the current neuromuscular condition and guide stimulation adjustment. In practice, the \textit{fresh} state allows stimulation to be maintained or slightly increased, the \textit{transitional} state suggests adjustment of stimulation intensity, and the \textit{fatigued} state indicates that stimulation should be reduced or terminated to prevent excessive muscular loading.

These results demonstrate that stimulation-evoked MMG signals contain sufficient information to reliably identify fatigue states using lightweight machine learning classifiers. The proposed framework therefore enables real-time fatigue estimation within the MMG–FES system and provides a practical decision mechanism for adaptive electrostimulation control.

\subsection{Closed-Loop System Behavior and Muscle Response}

To evaluate the therapeutic effectiveness of electrical stimulation massage and to verify the concurrent validity of the proposed MMG-FES sensing framework, a repeated stimulation–exercise protocol was conducted under a fatigued muscle state. Participants first performed squat exercises targeting the quadriceps femoris until a clear fatigue sensation was reported. Electrical stimulation was then applied for approximately one minute, followed by a subsequent squat set (10 repetitions). This stimulation–exercise cycle was repeated multiple times to examine how muscle mechanical state evolved across successive stimulation sessions.

\begin{figure}[!htbp]
    \centering
    \includegraphics[width=0.9\linewidth]{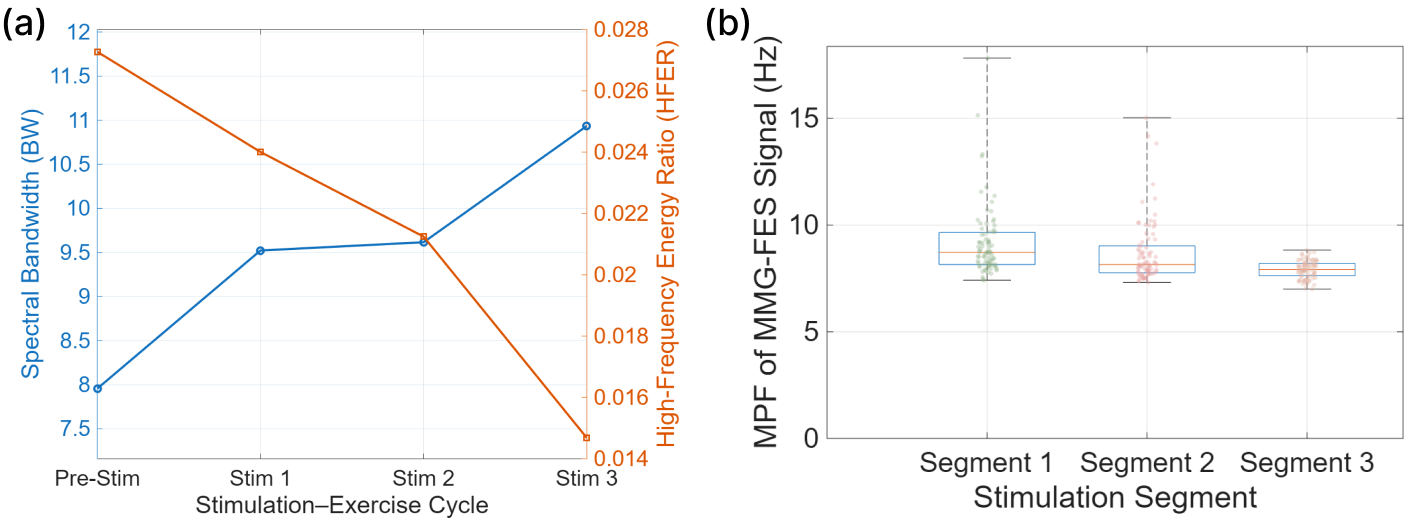}
    \caption{MMG-FES–based validation of electrical stimulation–induced muscle relaxation. (a) BW increases and HFER decreases across stimulation–exercise cycles during squat movements. (b) RMS amplitude during stimulation increases across segments, indicating progressively enhanced mechanical responsiveness.}
    \label{test1}
\end{figure}

As shown in Figure.\ref{test1} (a), MMG-FES features extracted from squat movements exhibited a systematic and monotonic trend across stimulation–exercise cycles. Specifically, spectral bandwidth (BW) increased progressively from the pre-stimulation condition to later cycles, while the high-frequency energy ratio (HFER) showed a corresponding decrease. According to the fatigue–relaxation relationships established previously, this combined increase in BW and reduction in HFER indicates a gradual transition from a fatigued, mechanically constrained state toward a more relaxed muscle condition following repeated electrical stimulation.

Complementary evidence is provided by the stimulation-phase MPF distribution shown in Fig.~\ref{test1}(b). Based on our prior validation, a decrease in MPF reflects a relaxation-dominant mechanical response with reduced fatigue-related high-frequency components, whereas an increase indicates fatigue accumulation.

\begin{table}[!t]
\centering
\caption{MPF statistics across stimulation segments}
\label{tab:mpf_segments}
\begin{tabular}{lccc}
\hline
Segment & Mean $\pm$ SD (Hz) & Median (Hz) & N \\
\hline
Segment 1 & 9.17 $\pm$ 1.66 & 8.72 & 83 \\
Segment 2 & 8.70 $\pm$ 1.48 & 8.15 & 83 \\
Segment 3 & 7.92 $\pm$ 0.42 & 7.91 & 83 \\
\hline
\end{tabular}
\end{table}

 As summarized in Table~\ref{tab:mpf_segments}, MPF decreased progressively across stimulation segments. The mean MPF declined from 9.17 $\pm$ 1.66 Hz (Segment 1) to 8.70 $\pm$ 1.48 Hz (Segment 2) and further to 7.92 $\pm$ 0.42 Hz (Segment 3). 
Median values showed a consistent downward trend (8.72 → 8.15 → 7.91 Hz). 
A Kruskal–Wallis test confirmed significant differences among segments ($p = 2.21 \times 10^{-12}$), with all pairwise Dunn–Šidák comparisons remaining significant ($p < 0.01$).

The monotonic reduction in MPF, together with the narrowing dispersion observed in Segment 3, indicates a stabilization of mechanical vibration patterns under continued stimulation, consistent with a cumulative relaxation effect.

\section{Discussion}
\begin{figure}[!htbp]
    \centering
    \includegraphics[width=0.9\linewidth]{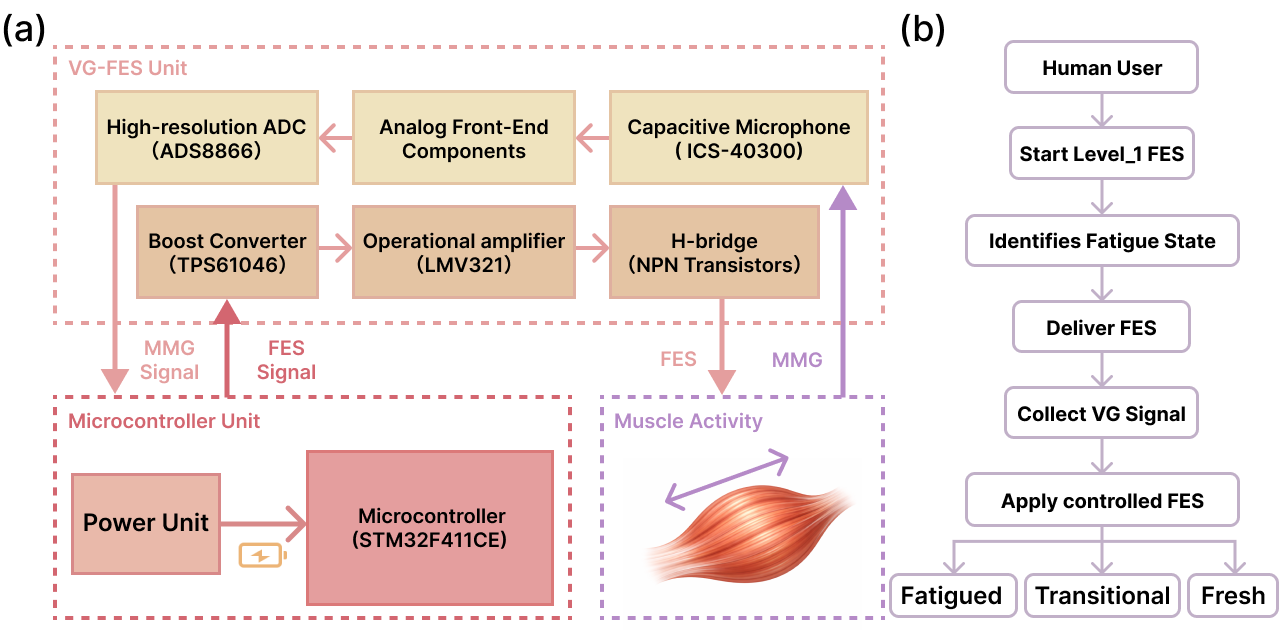}
    \caption{Human-in-the-loop MMG-FES system. (a) Closed-loop interaction between vibration-based fatigue identification and adaptive electrical stimulation. (b) Human-in-the-loop control flow: probing stimulation, fatigue classification, graded FES delivery, and RMS-based recovery monitoring.}
    \label{sys}
\end{figure}

The proposed system establishes a human-in-the-loop closed-loop framework that integrates fatigue estimation with adaptive electrical stimulation, as illustrated in Fig.~\ref{sys}. Unlike conventional FES systems that separate sensing and actuation or rely on open-loop parameter tuning, the proposed framework directly leverages stimulation-evoked MMG responses to simultaneously assess muscle condition and regulate stimulation. This unified design enables continuous neuromuscular monitoring during active stimulation and eliminates the need for alternating sensing–stimulation phases.

A key advantage of this approach is that fatigue assessment no longer depends on voluntary movement. By applying probing stimulation and analyzing the resulting mechanical responses, the system can infer fatigue states directly under stimulation-on conditions. This is particularly important in rehabilitation scenarios where voluntary control is limited or inconsistent. In addition, compared to EMG-based approaches, which are severely affected by stimulation artifacts, MMG signals remain stable and interpretable during FES. The observed consistency between MMG-derived spectral features and established EMG fatigue trends further supports the physiological relevance of the proposed method, suggesting that MMG captures meaningful neuromuscular changes rather than incidental vibration responses.

From a system perspective, the integration of fatigue estimation and stimulation control provides a practical pathway toward fatigue-aware closed-loop neuromuscular stimulation. Instead of applying fixed stimulation parameters, the system can adapt stimulation intensity and timing according to the detected muscle state, potentially reducing excessive fatigue accumulation and improving stimulation efficiency. Compared to existing approaches that rely on offline analysis or artifact-prone sensing modalities, the proposed framework offers a more robust and real-time solution for neuromuscular monitoring and control.

Despite these advantages, several limitations should be noted. First, the study was conducted on a limited number of healthy participants, and further validation on clinical populations, such as stroke or spinal cord injury patients, is required. Second, the proposed fatigue representation is primarily based on spectral features, and its robustness across different movement types, contraction intensities, and inter-subject variability remains to be further investigated. Finally, incorporating quantitative biomechanical measurements, such as force or torque, could further strengthen the physiological interpretation of MMG signals. Future work will focus on validating the system in real-world rehabilitation scenarios and integrating it into fully wearable, closed-loop stimulation platforms.

\section{Conclusion}
This study presents an MMG-based sensing and stimulation framework for fatigue-aware neuromuscular monitoring during functional electrical stimulation (FES). By integrating MEMS microphone–based mechanomyography (MMG) sensing with an FES wearable system, the proposed approach enables stable acquisition of muscle vibration signals under active stimulation, overcoming the limitations of conventional EMG measurements.

Experimental results demonstrate that MMG-derived spectral features, including MPF, BW, and HFER, effectively capture fatigue-related mechanical characteristics of muscle activity. Through a cross-condition learning strategy, fatigue states characterized during voluntary movement can be reliably recognized from stimulation-evoked MMG signals, enabling fatigue assessment in stimulation-on conditions.

These findings establish MMG as a physiologically meaningful and stimulation-compatible sensing modality and support its use in fatigue-aware closed-loop neuromuscular stimulation. The proposed framework provides a foundation for future adaptive rehabilitation systems that integrate real-time sensing and stimulation to improve therapeutic outcomes.

 
\bibliographystyle{IEEEtran}
\bibliography{bibs}

\end{document}